\RequirePackage{fix-cm}
\documentclass[twocolumn]{svjour3}          
\smartqed  
\usepackage{graphicx}

\usepackage[utf8]{inputenc} 
\usepackage[T1]{fontenc}    
\usepackage{hyperref}       
\usepackage{url}            
\usepackage{booktabs}       
\usepackage{amsfonts}       
\usepackage{nicefrac}       
\usepackage{microtype}      
\usepackage{cite}
\usepackage{subfigure}
\usepackage{booktabs}
\usepackage{amsfonts,bbm,mathrsfs}
\usepackage{amsmath}
\usepackage{algorithm,algorithmic,bm,color}
\usepackage{multirow}
\usepackage{wrapfig}
\usepackage{adjustbox}
\usepackage{graphics, color,  dsfont}

\usepackage{graphicx,psfrag,dsfont}
\usepackage{amsfonts,amsmath,amssymb,color}
\usepackage{algorithm,algorithmic,bm,color}
\usepackage{bbm,setspace}
\usepackage[flushleft]{threeparttable} 
\usepackage{color}
\usepackage[table]{xcolor} 
\usepackage{multirow}
\newcommand{\argmin}{\arg\!\min}

\newcommand{\Cmat}{{\bf C}}
\newcommand{\Dmat}{{\bf D}}

\newcommand{\Hmat}[0]{{{\bf H}}}
\newcommand{\Imat}{{\bf I}}

\newcommand{\Rmat}[0]{{{\bf R}}}

\newcommand{\Tmat}[0]{{{\bf T}}}

\newcommand{\Xmat}{{\bf X}}
\newcommand{\Ymat}[0]{{{\bf Y}}}
\newcommand{\Zmat}{{\bf Z}}

\newcommand{\pv}[0]{{\boldsymbol{p}}}
\newcommand{\qv}[0]{{\boldsymbol{q}}}

\newcommand{\uv}[0]{{\boldsymbol{u}}}
\newcommand{\vv}{\boldsymbol{v}}
\newcommand{\wv}{\boldsymbol{w}}

\newcommand{\xv}{\boldsymbol{x}}
\newcommand{\yv}{\boldsymbol{y}}

\newcommand{\zv}{\boldsymbol{z}}

\newcommand{\ts}{^{\top}}
\newcommand{\ie}{{\em i.e.}}
\newcommand{\eg}{{\em e.g.}}

\newcommand{\inv}{^{-1}}

\begin{document}

\title{Adaptive Deep PnP Algorithm for Video Snapshot Compressive Imaging}

\titlerunning{Adaptive PnP for Video SCI}        

\author{Zongliang Wu \and Chengshuai Yang \and Xiongfei Su \and Xin Yuan}


\authorrunning{Wu, Yang, Su and Yuan} 

\institute{The authors are at School of Engineering, Westlake University, Hangzhou, Zhejiang 310024, China\\
              \email \{wuzongliang, yangchengshuai, suxiongfei, xyuan\}@westlake.edu.cn \\
\protect
 Z. Wu and C. Yang contribute equally to this paper.
 \\ \protect
Corresponding author:  Xin Yuan
}

\date{ }

\maketitle
\begin{abstract}
Video Snapshot compressive imaging (SCI) is a promising technique to capture high-speed videos, which transforms the imaging speed from the detector to mask modulating and only needs a single measurement to capture multiple frames. The algorithm to reconstruct high-speed frames from the measurement plays a vital role in SCI. In this paper, we consider the promising reconstruction algorithm framework, namely plug-and-play (PnP), which is flexible to the encoding process comparing with other deep learning networks. 
One drawback of existing PnP algorithms is that they use a pre-trained denoising network as a plugged prior while the training data of the network might be different from the task in real applications. 
Towards this end, in this work, we propose the {\bf\em online PnP} algorithm  which can adaptively update the network's parameters within the PnP iteration; this makes the denoising network more applicable to the desired data in the SCI reconstruction. 
Furthermore, for color video imaging, RGB frames need to be recovered from Bayer pattern or named demosaicing in the camera pipeline. 
To address this challenge, we design a two-stage reconstruction framework to optimize these two coupled ill-posed problems and introduce a deep demosaicing prior specifically for video demosaicing. 
Extensive results on both simulation and real datasets verify the superiority of our adaptive deep PnP algorithm.
{The code to reproduce the results is at \url{https://github.com/xyvirtualgroup/AdaptivePnP_SCI}}.
%
\end{abstract}


\maketitle


\section{Introduction \label{Sec:intro}}
Video snapshot compressive imaging (SCI) \cite{Llull:13_y, 2014CVPR_xin_y} has attracted much attention because it can improve the imaging speed by capturing 3D information from one 2D measurement. When video SCI works, multiple frames are first modulated by different masks, and these modulated frames are mapped into a single measurement. After this, the reconstruction algorithm is employed to recover these multiple frames from the single measurement~\cite{Yuan2021_SPM_y,jalali2019snapshot_performace}. At present, the masks can easily be adjusted with a higher speed than the capture rate of the camera~\cite{Hitomi11ICCV_y,Qiao2020_APLP_y,2011CVPR_Reddy_y}. Thus, SCI enjoys the advantages of high speed, low memory, low bandwidth, low power and potentially low cost~\cite{Yuan_2020_CVPR_y}. 


Due to the fact that the final (desired high-speed) frames are obtained by reconstruction algorithms, various algorithms have been developed during past few years to improve the performance of SCI. These algorithms lie in two 
categories: optimization-based and deep learning-based methods. Among the optimization ones, TwIST~\cite{Bioucas-Dias2007TwIST_y}, Gaussian Mixture Model (GMM) in~\cite{Yang14GMMonline_y,Yang14GMM_y} and GAP-TV~\cite{Liao14GAP_y,Yuan16ICIP_GAP_y} have a high computing speed but could not achieve high quality of images, while DeSCI~\cite{Liu18TPAMI_y} can obtain high quality images, but the computing process 
takes a long time. 
Recently, many deep learning methods have been developed to reconstruct the videos for the SCI system~\cite{Qiao2020_APLP_y,2020EndLi_y,conf/eccv/ChengLWZCMY20_y,Cheng2021_CVPR_ReverSCI_y,song2021memory}.  For typical ill-posed problem of SCI, optimization inspired physics-driven networks (deep unfolding) \cite{Ma19ICCV_y,Meng_GAPnet_arxiv2020,wu2021DUN-3DUnet} have been proposed. Though these methods can finish the task within seconds (after training) and achieve the state-of-the-art results, they lose the robustness of the network since whenever the sensing matrix (encoding process) changes, a new network has to be re-trained, or fine tuned~\cite{Wang2021_CVPR_MetaSCI_y}. 
This makes them challenging for adaptive video sensing~\cite{Yuan13ICIP_y}.

\begin{figure*}[t!] 
\centering
\includegraphics[width=1\linewidth]{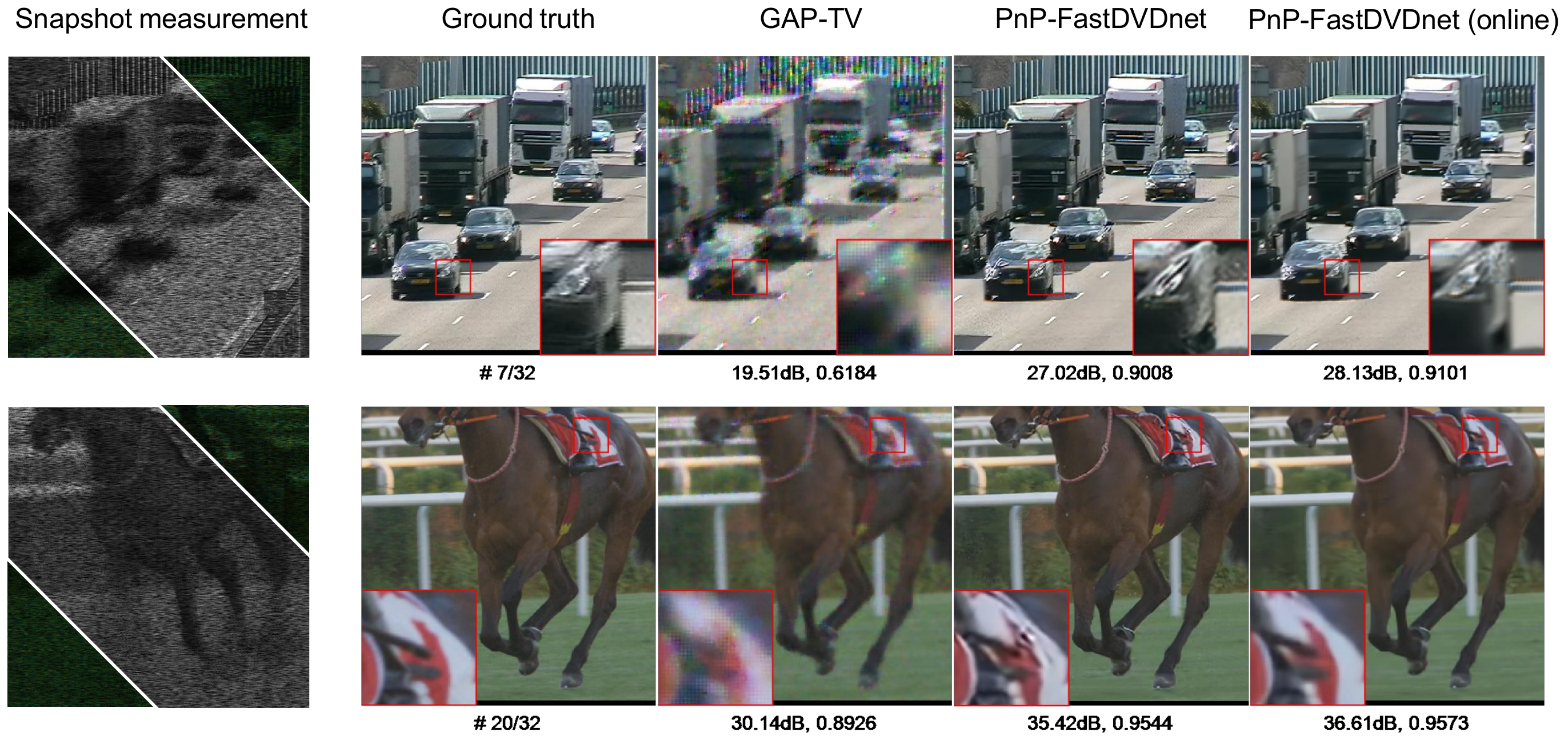}
\vspace{-4mm}
\caption{Reconstructed mid-scale color SCI images of two scenes, \texttt{Traffic} (upper) and \texttt{Jockey} (lower). One out of 8 reconstructed frames from a snapshot measurement (left) is shown for each scene, comparing our proposed {\bf PnP-FastDVDNet (online)} with the previous state-of-the-art PnP-FastDVDNet~\cite{PnP_SCI_TPAMI2021} and GAP-TV~\cite{Yuan16ICIP_GAP_y} as a baseline. 
}
\label{fig:mid_tease}
\end{figure*}

\begin{figure*}[!htbp]
	\begin{center}
        \includegraphics[width=1\linewidth]{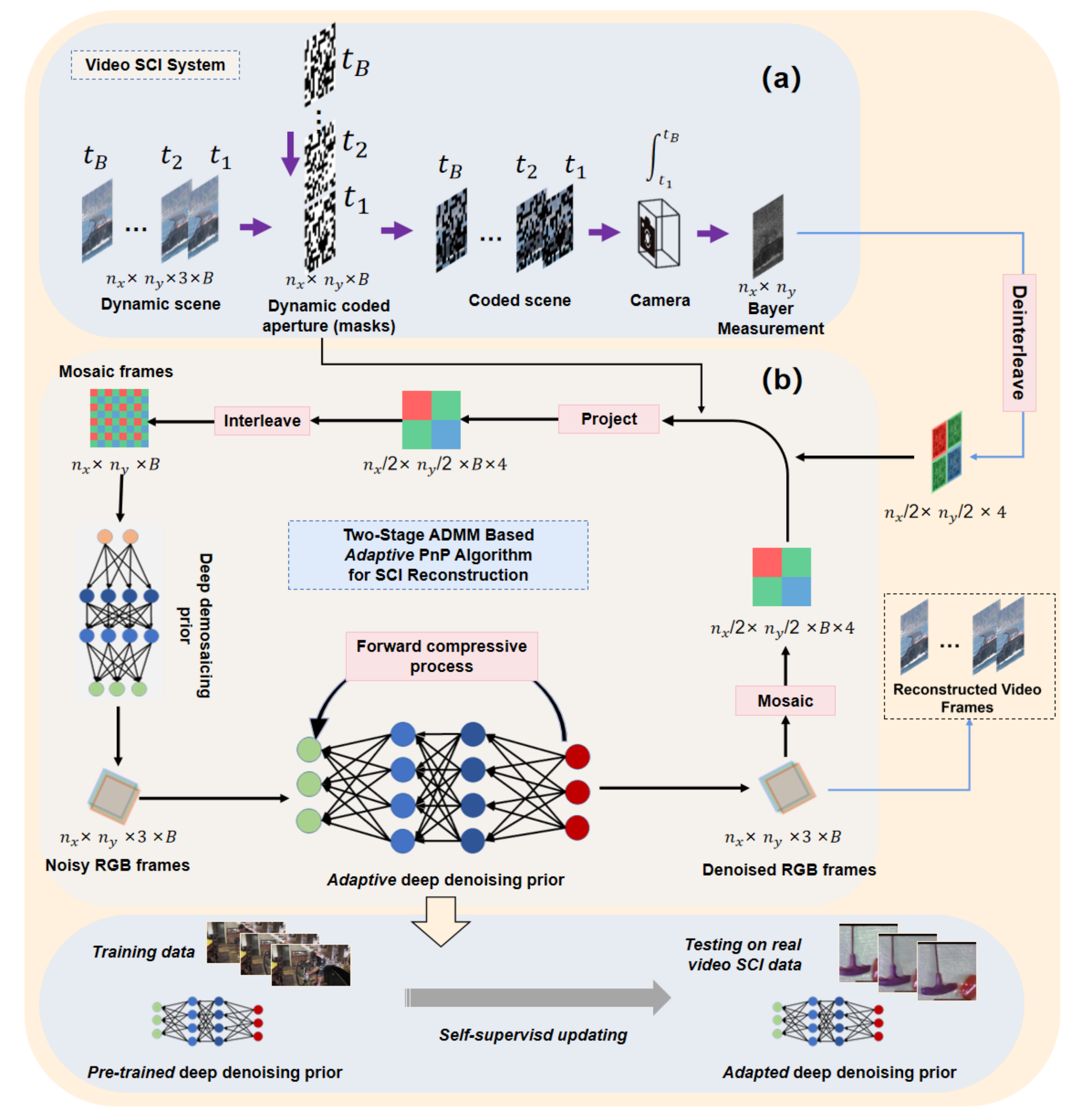}
	\end{center}
	\vspace{-3mm}
	\caption{ (a) Video SCI system: the high speed dynamic scene, here shown as the color RGB image sequences at timestamps $t_1$, $t_2$ to $t_B$, are encoded by the high-speed variant masks (dynamic coded apertures); these coded scenes are captured by a color Bayer pattern camera within one exposure time $\int_{t_1}^{t_B}$ to form a single 2D Bayer measurement (top-right). (b) The captured Bayer measurement and the pre-defined (or calibrated) mask patterns are sent to the proposed {\em adaptive} Plug-and-Play (PnP) algorithms to reconstruct the desired high-speed color video (bottom-right).
	Our proposed adaptive PnP includes two parts:  the newly trained demosaicing network (bottom-left) and the {\em adaptive} deep denoising prior, in which the parameters will be adapted during the reconstruction process to better fit the desired data.}
	\label{fig:principle}
	\vspace{-2mm}
\end{figure*}

Fortunately, the Plug-and-Play (PnP) framework \cite{Venkatakrishnan_13PnP_y,Sreehari16PnP_y,Chan2017PlugandPlayAF_y,2019PlugRyu_y} provides a feasible way to develop an efficient and flexible algorithm for video SCI reconstruction, especially for large-scale problems under the framework of alternating direction method of multipliers (ADMM)~\cite{Boyd11ADMM_y} or generalized alternating projection (GAP)~\cite{Liao14GAP_y}. Therefore, this paper focuses on the PnP algorithm to upgrade video SCI performance.
Specifically, we propose an {\em adaptive PnP} algorithm that updates the denoising network in the PnP iteration so that the network can better fit the current measurement (Fig.~\ref{fig:principle}). We dub the proposed algorithm {\bf online PnP}. 
Besides, for more realistic applications of color video imaging, we proposed a {\em Two-Stage ADMM} framework and a deep video demosacing prior to solve this two-stage ill-posed problem.

\subsection{Motivation}
Even though the PnP based on the {\em pre-trained} FFDNet~\cite{Yuan_2020_CVPR_y} can achieve an acceptable result  for video SCI. There is some room for improvement. Firstly, the FFDNet~\cite{Zhang18TIP_FFDNet_y} aims at denoising a single image instead of video while the SCI reconstruction aims at a series of images in one video. Secondly, the pre-trained parameters in one network may not match the specific testing dynamic scene, especially the one captured by the real system. Thirdly, the color video may demand a different PnP framework to recover. Most recently, Yuan et al.~\cite{PnP_SCI_TPAMI2021} have extended the PnP-FFDNet and PnP-FastDVDNet by using the recent state-of-the-art video denoiser, FastDVDNet~\cite{Tassano_2020_CVPR_y} to achieve better results, especially on the color Bayer video SCI. 
However, there are still two research gaps desired to fill: 
\begin{itemize}
    \item [$i$)] the mismatch between the pre-trained denoiser (FFDNet or FastDVDNet) and the targeted video in SCI, and
    \item [$ii$)] the PnP-color in \cite{Yuan_2020_CVPR_y} usually achieves better results for color videos but still produces color artifacts and the demosaicing algorithms used in~\cite{PnP_SCI_TPAMI2021} based on optimization are more than 15 years old~\cite{malvar2004high-quality_y,Menon07_y}.
\end{itemize}
Recent researches have shown promising results of using deep learning based demosaicing~\cite{Brady:20_y,GharbiACM16_y}, and therefore, the color video SCI reconstruction based on PnP should lead to better results using a deep demosaicing method.

\subsection{Contributions}
Bearing the above concerns in mind, we aim to further improve the performance of the PnP algorithm for video SCI.
Specific contributions of this paper are listed as follows:
\begin{itemize}
\item Considering the fact the pre-trained network parameters may not be best for one specific testing scene, we develop the {\em adaptive PnP} method by automatically updating the parameters, such as weights and bias, in the network of deep denoising prior according to the specific dynamic scene.  

\item In order to better apply PnP to color video, we develop a new {\em Two-Stage ADMM} optimization regime to solve the two stage ill-posed problem, \ie, the SCI reconstruction and demosaicing process.

\item We introduce a Deep Demosaicing Network (DDNet) as prior especially for videos into our proposed method.
Please refer to Fig.~\ref{fig:mid_tease} as an example, where we can see that more than 1dB in PSNR improvement has been achieved.

\item We consider sequential measurements in video SCI, \ie,  using the network parameters updated from the previous measurement to initialize the network used in the next measurement. This saves computational complexity and thus saves running time.

\item We successfully apply the adaptive denoising method and our deep demosaicing model to color video SCI and have achieved better results than the PnP-FastDVDNet developed in~\cite{PnP_SCI_TPAMI2021}. Extensive results on both simulation and real datasets verify the superiority of our adaptive deep PnP algorithm. 
\end{itemize}

\subsection{Related Work and Organization of This Paper}
In fact, PnP framework belongs to the optimization methods, which can date back to 2013~\cite{Venkatakrishnan_13PnP_y} and used in image restoration\cite{zhang2021plug}. Due to the fact that PnP is easy to combine a variety of denoising priors, it becomes more powerful than before, with the emergence of advanced deep denoising networks~\cite{Zhang17SPM_deepdenoise_y,Zhang18TIP_FFDNet_y}. Initially, only DeSCI~\cite{Liu18TPAMI_y} among the optimization methods can obtain high quality of images reconstructed from a single measurement in SCI. However, the slow speed of DeSCI precludes its real applications. Even though deep learning methods~\cite{Ma19ICCV_y,Qiao2020_APLP_y,2020EndLi_y,conf/eccv/ChengLWZCMY20_y,Wang2021_CVPR_MetaSCI_y,Cheng2021_CVPR_ReverSCI_y,wu2021DUN-3DUnet} can achieve the state-of-the-art results within seconds (after training), they lose the robustness of the network for new sensing matrix. By contrast, the previous study~\cite{Yuan_2020_CVPR_y} has proved that PnP combining deep denoising prior can provide an efficient and flexible method for video SCI reconstruction. Besides, the self-supervised implementation of putting deep image prior\cite{ulyanov2018dip} into PnP framework~\cite{meng2021pnpdip} can reach a great performance even better than supervised algorithms for the similar task for spectral SCI reconstruction.
In this paper, we aim to explore how to improve the deep PnP based on the {\em adaptive deep prior}. 

The rest of this paper is organized as follows: Section~\ref{Sec:SCI_model} reviews the mathematical model of SCI and introduces the color SCI forward model with a Bayer pattern sensor. Section~\ref{Sec:PnP_SCI} describes the PnP-SCI algorithms for color video SCI using denoising networks and demosaicing networks based on ADMM. 
The proposed adaptive PnP is presented in Section~\ref{Sec:PnP_online} with details of the proposed deep demosaicing network introduced in Section~\ref{Sec:PnP_SCI}.
Extensive results on both simulation data, including the gray-scale benchmark, color Bayer mid-scale benchmark, and real data are presented in Section~\ref{Sec:Results}.
Section~\ref{Sec:conclu} concludes the entire paper.

\section{Preliminary: Video SCI System \label{Sec:SCI_model}}
As mentioned in the introduction, the key idea of video SCI is to use different masks to modulate video frames at different timestamps. As depicted in Fig.~\ref{fig:principle} (a), let $\{\Xmat_1, \dots, \Xmat_{B}\}$ denote the discretized video frames, either color or gray-scale, at timestamps $\{t_1, \dots, t_B\}$. These video frames are modulated by dynamic coded aperture, a.k.a., the masks $\{\Cmat_1, \dots, \Cmat_B\}$, respectively. These modulated frames are then integrated to a single coded measurement (a compressed image).  It is easy to understand that the key component of SCI system is the modulation device, which needs to be working at a higher frequency than the capture rate of the camera being used. In the literature, widely used devices are the shifting mask~\cite{Patrick13OE_y,2014CVPR_xin_y} and digital micromirror devices~\cite{2011CVPR_Reddy_y,Qiao2020_APLP_y}. 

While it is straightforward to write down the forward model of grayscale video SCI, it might be non-trivial for the color SCI case, since a Bayer pattern sensor has spatially variant ``RGGB" channels ~\cite{2014CVPR_xin_y}.
This makes the color video SCI actually include two steps.
i) The mask modulates all the RGB (R--Red, G--Green, B--Blue) channels in the desired video frames.
ii) After the modulation and integration, the ``R'', ``G1'', ``G2'' and ``B'' pixels on the sensor conduct a down-sampling of the compressed measurement to form the final Bayer snapshot measurement. 

To be concrete, consider the desired RGB color video is $\{\Xmat_b\}_{b=1}^B \in {\mathbb R}^{n_x\times n_y\times 3}$. The final compressed measurement is  $\Ymat\in {\mathbb R}^{n_x\times n_y}$. 
Regarding the Bayer pattern, the raw measurement (top-right in Fig.~\ref{fig:principle} (a)) is decoupled into four components $\{\Ymat^{(r)},\Ymat^{(g_1)},\Ymat^{(g_2)},\Ymat^{(b)}\} \in {\mathbb R}^{\frac{n_x}{2}\times \frac{n_y}{2}}$. Similarly, the corresponding masks and videos are denoted by $\{\Cmat^{(r)},\Cmat^{(g_1)},\Cmat^{(g_2)},\Cmat^{(b)}\} \in {\mathbb R}^{\frac{n_x}{2}\times \frac{n_y}{2}\times B}$, $\{\tilde{\Xmat}^{(r)},\tilde{\Xmat}^{(g_1)},\tilde{\Xmat}^{(g_2)},\tilde{\Xmat}^{(b)}\} \in {\mathbb R}^{\frac{n_x}{2}\times \frac{n_y}{2}\times B}$, respectively. 
The forward model for each channel is now
\begin{equation}
\Ymat^{(i)} =  \sum_{b=1}^B \Cmat^{(i)}_b\odot \tilde{\Xmat}^{(i)}_b + \Zmat^{(i)}, \quad \forall i = \{r, g_1, g_2,b\},
\end{equation}
where $\Zmat$ denotes the noise term for each channel.
The demosaicing is basically an interpolation process from $\tilde{\Xmat}^{(r)}$ to $\Xmat^{(r)} \in \mathbb {R}^{n_x\times n_y\times B}$, from $\{\tilde{\Xmat}^{(g_1)},\tilde{\Xmat}^{(g_2)}\}$  to ${\Xmat}^{(g)} \in  \mathbb {R}^{n_x\times n_y\times B}$ and from $\tilde{\Xmat}^{(b)}$  to ${\Xmat}^{(b)} \in  \mathbb {R}^{n_x\times n_y\times B}$. It should be noted that the interpolation rate for Red and Blue channel is from 1 pixel to 4 pixels, whereas for the Green channel it is from 2 pixels to 4 pixels.

In the following, we use $\Tmat$ to denote the mosaicking process, \ie, from an RGB image to a Bayer pattern mosaicked RGGB image. In a vectorized formulation, let $\xv_b \in {\mathbb R}^{3n_xn_y}$ denote the RGB image and $\tilde{\xv}_b \in {\mathbb R}^{n_xn_y}$ denote the mosaicked RGGB image; we have 
\begin{equation}
    \tilde{\xv}_b = \Tmat \xv_b, \quad \forall b = 1,\dots,B, \label{Eq:mosaic}
\end{equation}
where $\Tmat\in {\mathbb R}^{n_xn_y \times 3n_xn_y}$. In color video SCI, from $\{\tilde{\xv}_b\}_{b=1}^{B}$ to the measurement $\yv \in {\mathbb R}^{n_xn_y}$ is the same as the gray-scale video SCI, which has been derived before such as in~\cite{Yuan_2020_CVPR_y}. This is
\begin{eqnarray}
    \yv &= &\Hmat  \tilde{\xv} + \zv,  \label{Eq:forward_1} \\
    \tilde{\xv}&=& [\tilde{\xv}_b\ts, \dots, \tilde{\xv}_B\ts]\ts, \\
    \Hmat &=& [\Dmat_1, \dots, \Dmat_B],  \label{Eq:Hmat}
\end{eqnarray}
where $\Dmat_b  = {\rm Diag}({\rm Vec}(\Cmat_b))$, with ${\rm Vec}(\cdot)$ vectoring the ensued matrix and ${\rm Diag}(\cdot)$ diagonalizing the vector within the $(~)$ and $\Hmat\in {\mathbb R}^{n_xn_y \times B n_xn_y}$ is thus the sensing matrix; $\zv\in {\mathbb R}^{n_xn_y}$ denotes the vectorized noise.

Plug \eqref{Eq:mosaic} into \eqref{Eq:forward_1}, we have
\begin{eqnarray}
    \yv &=& \Hmat \left[\begin{array}{c}
         \Tmat \xv_1\\
         \vdots\\
         \Tmat \xv_B
    \end{array}\right] + \zv\\
    &=& \Hmat \underbrace{\left[ \begin{array}{cccc}
    \Tmat & {\bf 0} & \cdots & {\bf 0}\\
    {\bf 0} & \Tmat &  \cdots & {\bf 0}\\
    \vdots & \vdots  & \ddots & \vdots\\
    {\bf 0}  & {\bf 0}   & \cdots  & \Tmat\\
        \end{array}\right]}_{\stackrel{\rm def}{=} \Tmat_{\cal M}} 
        \underbrace{\left[ \begin{array}{c}
        \xv_1 \\
        \vdots \\
        \xv_B
        \end{array}\right]}_{\stackrel{\rm def}{=}\xv} + \zv \\
        &=& \Hmat \Tmat_{\cal M} \xv + \zv, \label{eq:forward_one}
\end{eqnarray}
where $\Tmat_{\cal M} \in {\mathbb R}^{Bn_xn_y \times 3B n_xn_y}$ can be recognized as the video mosaicking operation, \ie, imposing mosaicking on each frame. 
Note that for the gray-scale videos, $\xv \in {\mathbb R}^{B n_xn_y}$ and $\Tmat_{\cal M}$ will be an identity matrix.

Based on the model derived in \eqref{eq:forward_one}, in the following, we develop the adaptive deep  method for color video SCI reconstruction using the PnP framework.

\section{PnP Algorithm for Color Video SCI~\label{Sec:PnP_SCI}}
Different from the algorithm proposed in~\cite{Yuan_2020_CVPR_y}, in the following, we drive an end-to-end solution for color video SCI using the ADMM framework.  
From \eqref{eq:forward_one}, we can see that given $\yv, \Hmat$ and $\Tmat_{\cal M}$, solving $\xv$ is an ill-posed problem. In fact, a {\em two-stage ill-posed problem, where one stage is from $\Tmat_{\cal M}$ and the other stage comes from $\Hmat$}.  As being widely used in other inverse problem, a prior $g(\xv)$ is employed, and thus the inverse problem is modeled as
\begin{equation}
    \hat{\xv} = \argmin_{\xv} \frac{1}{2}\|\yv - \Hmat \Tmat_{\cal M} \xv\|_2^2 + \lambda g(\xv),
\end{equation}
where $\lambda$ is a parameter to balance the fidelity term and the prior. 
Different from conventional algorithms, the key innovation of PnP is introducing the implicit prior, $g(\xv)$, where diverse priors including the deep neural network~\cite{Zhang17SPM_deepdenoise_y} based one can be plugged into the algorithm.

As mentioned in~\cite{PnP_SCI_TPAMI2021}, a simple solution is to treat $\Hmat \Tmat_{\cal M}$ jointly and invoke ADMM to solve the problem. However, due to the challenges in SCI (temporal CS) and demosaicing (spatial interpolation), this joint modeling cannot lead to good results.
To address this challenge, we hereby propose a two-stage PnP-ADMM framework.

\subsection{Two-Stage PnP-ADMM}
We firstly decouple the video SCI and demosaicing process by introducing a parameter $\qv$:
\begin{equation}
     \hat{\xv} = \argmin_{\xv} \frac{1}{2}\|\yv - \Hmat \qv \|_2^2 + \lambda g(\xv),~ {\text {subject to}}~\qv = \Tmat_{\cal M} \xv.
\end{equation}
In the following, we introduce a parameter $\uv$ to be updated and another $\rho$ to be manually set, and get the augmented Lagrangian function
\begin{equation}
    \begin{aligned}
    {\cal L}_{\rho}(\qv, \xv, \uv) =& \frac{1}{2}\|\yv - \Hmat \qv \|_2^2 + \lambda g(\xv) \\
    &+ \frac{\rho}{2}\|\qv - \Tmat_{\cal M} \xv\|_2^2 + \uv\ts(\qv - \Tmat_{\cal M} \xv). 
    \end{aligned}
    \label{Eq:L_function}
\end{equation}
Using ADMM, the minimization of ${\cal L}$ in \eqref{Eq:L_function} can be split into the following three sub-problems:
\begin{align}
    \hat{\qv}& = \argmin_{\qv} \frac{1}{2}\|\yv - \Hmat \qv \|_2^2  + \frac{\rho}{2}\|\qv - \Tmat_{\cal M} \xv|_2^2   \nonumber\\ & \qquad \qquad\qquad + \uv\ts(\qv - \Tmat_{\cal M} \xv), \nonumber\\
    & = \argmin_{\qv} \frac{1}{2}\|\yv - \Hmat \qv \|_2^2  + \frac{\rho}{2}\left\|\qv  - (\Tmat_{\cal M} \xv - \frac{1}{\rho}\uv)\right\|_2^2, \label{Eq:qv}\\
    \hat{\xv}& = \argmin_{\xv} \frac{\rho}{2}\|\Tmat_{\cal M} \xv - \qv\|_2^2 + \uv\ts(\qv - \Tmat_{\cal M} \xv) +\lambda g(\xv), \nonumber\\
    & =  \argmin_{\xv}\frac{\rho}{2} \left\|\Tmat_{\cal M} \xv - (\qv+ \frac{1}{\rho}\uv)\right\|_2^2  + \lambda g(\xv),  \label{Eq:xv}\\
    \hat{\uv} &= \argmin_{\uv} \uv\ts(\qv - \Tmat_{\cal M} \xv). \label{Eq:uv}
\end{align}
These three variables can be solved by fixing the other two with an iterative optimization. 
Let the superscript $k$ denotes the iteration number.
\subsubsection{sub-problem of ${\bf q}$}
Eq.~\eqref{Eq:qv} is a quadratic formulation, which has a closed-form solution:
\begin{equation}
    (\Hmat\ts\Hmat + \rho \Imat) \qv^{(k+1)} = \Hmat\ts\yv + \rho (\Tmat_{\cal M} \xv^{(k)} - \frac{1}{\rho}\uv^{(k)}) \label{Eq:qvclosed}
\end{equation}
As described in \eqref{Eq:Hmat}, $\Hmat$ is a concatenation of diagonal matrices and thus $\Hmat\Hmat\ts$ is a diagonal matrix, \ie,
\begin{align}
    \Rmat &= \Hmat\Hmat\ts = {\rm Diag}(r_1, \dots, r_n), \quad  n=n_xn_y \label{Eq:Rmat} \\
    r_j &= \sum_{b=1}^B c_{b,j}^2,  \quad \forall j = 1,\dots, n. 
\end{align}
Using the matrix inversion formula, 
\begin{align}
     (\Hmat\ts\Hmat + \rho \Imat)\inv = \rho\inv \Imat - \rho\inv \Hmat\ts(\Imat + \Hmat\Hmat\ts/\rho)\inv\Hmat (\rho\inv\Imat).
\end{align}
Eq.~\eqref{Eq:qvclosed} leads to 
\begin{align}
&\qv^{(k+1)}  =   (\Hmat\ts\Hmat + \rho \Imat)\inv \left[\Hmat\ts\yv + \rho (\Tmat_{\cal M} \xv^{(k)} - \frac{1}{\rho}\uv^{(k)}) \right] \nonumber\\
& = \textstyle \rho\inv \Hmat\ts\yv +\Tmat_{\cal M} \xv^{(k)} - \frac{1}{\rho}\uv^{(k)}   - \rho\inv\Hmat\ts\left(\Imat + \frac{\Rmat}{\rho}\right)\inv \nonumber\\
& \quad \times \left[\frac{\Rmat\yv}{\rho} + \Hmat(\Tmat_{\cal M} \xv^{(k)} - \frac{1}{\rho}\uv^{(k)})\right]. \label{Eq:qk+1}
\end{align}
From~\eqref{Eq:Rmat}, 
\begin{align}
    \left(\Imat + \frac{\Rmat}{\rho}\right)\inv &= {\rm Diag}\left[\frac{\rho}{\rho+ r_1},\dots,\frac{\rho}{\rho+ r_n}\right] \\
    \left(\Imat + \frac{\Rmat}{\rho}\right)\inv \frac{\Rmat}{\rho} & = {\rm Diag}\left[\frac{r_1}{\rho+ r_1},\dots,\frac{r_n}{\rho+ r_n}\right]
\end{align}
Plugging into \eqref{Eq:qk+1}, we have
\begin{align}
  \pv^{(k)} &= \Tmat_{\cal M} \xv^{(k)} - \frac{1}{\rho}\uv^{(k)},\\
 \qv^{(k+1)} & =   \pv^{(k)} + \Hmat\ts \left[\frac{\yv_1 - [\Hmat \pv^{(k)} ]_1}{\rho+ r_1},\dots, \frac{\yv_n - [\Hmat \pv^{(k)} ]_n}{\rho+ r_n}\right]\ts , \label{Eq:qv_k+1}
\end{align}
where $[\Hmat \pv^{(k)} ]_j, \forall j = 1,\dots, n$ denotes the $j$-th element of the vector inside $[~]$. 
This can be solved efficiently in one shot.

\subsubsection{sub-problem of ${\bf u}$}
According to \eqref{Eq:uv}, $\uv$ is updated by gradient descent
\begin{equation}
    \uv^{(k+1)} = \uv^{(k)} + (\qv^{(k)} - \Tmat_{\cal M} \xv^{(k)}). \label{Eq:uv_k+1}
\end{equation}

\subsubsection{sub-problem of ${\bf x}$}
The solution to \eqref{Eq:xv} is not straightforward. Therefore, we hereby employ ADMM again to solve it.
Introducing parameters $\vv$ and $\wv$, we have
\begin{align}
    \hat{\xv}& =  \argmin_{\xv}\frac{\rho}{2} \left\|\Tmat_{\cal M}\xv - (\qv+ \frac{1}{\rho}\uv)\right\|_2^2  + \lambda g(\vv) \nonumber\\
    &\qquad\qquad \qquad  ~{\text{subject to}}~~ \vv =  \xv. 
\end{align}
We aim to minimize the following loss function
\begin{equation}
    \begin{aligned}
    {\cal L}_{\tau}(\xv, \vv, \wv) &= \frac{\rho}{2} \left\|\Tmat_{\cal M}\xv - (\qv+ \frac{1}{\rho}\uv)\right\|_2^2  + \lambda g(\vv) \\
    &\quad + \frac{\tau}{2}\|\xv - \vv\|_2^2 + \wv\ts(\xv - \vv).
    \end{aligned}
\end{equation}
Similar to the ADMM process from \eqref{Eq:L_function} to \eqref{Eq:uv}, we have the following three sub-problems. 

\begin{itemize}
    \item [3.1)] $\xv$ sub-problem:
    \begin{equation}
    \begin{aligned}
        \xv^{(k+1)}&=\argmin_{\xv}\frac{\rho}{2} \left\|\Tmat_{\cal M}\xv - (\qv^{(k)}+ \frac{1}{\rho}\uv^{(k)})\right\|_2^2 \\
        &\qquad + \frac{\tau}{2}\|\xv - ( \vv^{(k)}-\frac{1}{\tau}\wv^{(k)})\|_2^2. 
        \end{aligned}
        \label{Eq:x_k+1_demo}
    \end{equation}
    This is a quadratic form and has the following closed-form solution:
    \begin{equation}
    \begin{aligned}
         \xv^{(k+1)}= &[\rho \Tmat_{\cal M}\ts \Tmat_{\cal M} + \tau \Imat]\inv \\
         & \times [\rho \qv^{(k)}+ \uv^{(k)} + \tau\vv^{(k)}+ \wv^{(k)}]. \label{Eq:xv_k+1}
         \end{aligned}
    \end{equation}
    In fact, this is a demosaicing problem, \ie, estimating the demosaicked video $\xv$ from the mosaicked video $\qv^{(k)}+ \frac{1}{\rho}\uv^{(k)}$.
    
    The second term in \eqref{Eq:x_k+1_demo}, $\frac{\tau}{2}\|\xv - ( \vv^{(k)}-\frac{1}{\tau}\wv^{(k)})\|_2^2$, plays the role of a prior. 
    In the literature, various priors have been employed for demosaicing. In this work, inspired by the PnP framework, we have trained a deep demosaicing network to solve this problem with details shown in Section~\ref{Subsec:deep_demosa}, \ie, using the deep neural networks to implicitly replace the prior. We denote the solution by 
    \begin{equation}
        \xv^{(k+1)} =  {\cal D}_{\cal M}(\qv^{(k)}+ \frac{1}{\rho}\uv^{(k)}), \label{Eq:deep_dema}
    \end{equation}
    where ${\cal D}_{\cal M}$ is the deep demosaicing network.
    
    {In the experiments, we further compare the proposed deep demosaicing method with the closed-form solution in Eq.~\eqref{Eq:xv_k+1}. The results in Table~\ref{Tab:results_midscale} (Two-Stage ADMM-FFDNet-CDF) show that the deep demosaicing network has superior performance than the closed-form one, which also needs to use early stop to reduce performance decline in iterations.}
    \item [3.2)] $\vv$ sub-problem:
    \begin{align}
        \vv^{(k+1)} = \argmin_{\vv}\frac{\tau}{2} &\|\vv - (\xv^{(k+1)}-\frac{1}{\tau}\wv^{(k)})\|_2^2 + \lambda g(\vv).
    \end{align}
    As shown in the PnP work, this is the denoising problem and by using the deep denoising network such as FFDNet and FastDVDNet, the solution can be denoted as
    \begin{equation}
        \vv^{(k+1)} = {\cal D}_{\sigma}(\xv^{(k+1)}-\frac{1}{\tau}\wv^{(k)}), \label{Eq:vvk+1}
    \end{equation}
    where $\sigma$ is the estimated noise level and related to $\lambda/\tau$.
    
    \item [3.3)] $\wv$ sub-problem:
    \begin{equation}
        \wv^{(k+1)} = \wv^{(k)} + (\xv^{(k+1)} - \vv^{(k+1)}). \label{Eq:wvk+1}
    \end{equation}
\end{itemize}
From the above derivation, we can see that this is a two-stage ADMM framework. 

We re-order the updating equations and summarize the entire algorithm in Algorithm~\ref{algo:PnP_Color_Video}.

\begin{algorithm}[!htbp]
	\caption{Two-Stage PnP-ADMM for Color Video SCI}
	\begin{algorithmic}[1]
		\REQUIRE $\Hmat$, $\yv$, $\Tmat_{\cal M}$.
		\STATE Initial $\xv,\uv,\vv,\wv $ and $\rho, \tau$.
		\WHILE{Not Converge}
		\STATE Update $\qv$ by Eq.~\eqref{Eq:qv_k+1}. 
		\STATE Update $\xv$ by Eq.~\eqref{Eq:xv_k+1} or by the deep demosaicing network \eqref{Eq:deep_dema}.
		\STATE Update $\vv$ by deep denoiser in Eq.~\eqref{Eq:vvk+1}.
		\STATE Update $\wv$ by Eq.~\eqref{Eq:wvk+1}.
		\STATE Update $\uv$ by Eq.~\eqref{Eq:uv_k+1}.
		\ENDWHILE
	\end{algorithmic}
	\label{algo:PnP_Color_Video}
\end{algorithm}

\subsection{Deep Video Demosaicing Prior for PnP \label{Subsec:deep_demosa}}
In a color camera pipeline, demosaicing is a necessary process to recover RGB image from Color Filter Array (CFA, usually a Bayer RGGB pattern~\cite{malvar2004high-quality_y}). \\
For color video SCI, the demosaicing process is significantly important as well. The traditional demosaicing methods such as 'Malvar04'~\cite{malvar2004high-quality_y} and `Menon07'~\cite{Menon07_y} used in \cite{PnP_SCI_TPAMI2021} can achieve relative good results but are usually slow. 
Since the deep learning based demosaicing algorithm proposed~\cite{gharbi2016deepdemosaci}, a number of deep demosaicing using non-local property, self-attentions are developed and obtain higher recovery quality\cite{kokkinos2018deepdemosaci,mou2021dynamicdemosaci,xing2021enddemosaci}. Thus, to improve the performance of this demosaicing process, we need a deep demosaicing prior for our PnP.\\
However, existing demosaicing methods are mostly developed for a single image, without considering the inter-frame information and usually can not achieve a considerable performance for video demosaicing in our framework. Therefore, we borrow the idea from FastDVDNet and use a similar network structure as our deep demosaicing network (DDNet). Specifically, we retrained the network by removing batch-norm layers. 

For training data, we used 800 high definition images in the DAIVS 2017 480p dataset~\cite{perazzi2016davis,pont2017davis} as our training and validation set. The input mosaic images (RGGB Bayer mosaic images with 3 channels) can be generated by downsampling RGB channels from the original ones. 
We divide them into $64\times64$ patches and randomly choose 64 each time to input the proposed deep demosaicing network. After training, we plug it into the proposed adaptive PnP algorithm. As shown in Table~\ref{Tab:results_midscale} (Two-Stage ADMM-FFDNet-DDNet), the performance of demosaicing is improved by 0.38dB averagely in simulated middle-scale SCI data. All the training and testing of our deep demosaicing  method are conducted on an Nvidia RTX 3090 GPU with an Intel Xeon Gold 6256 CPU.

\section{Proposed Adaptive PnP Algorithm ~\label{Sec:PnP_online}}
It has been proved in~\cite{Yuan_2020_CVPR_y} that the reconstruction error term depends on the bounded denoising algorithm. In other words, a better denoiser can provide a reconstruction result closer to the true signal. Some special pre-trained deep denoising priors, such as FFDNet and FastDVDNet,  can provide the high quality of images at a high computing speed. However, for one specific dynamic scene, the pre-trained priors may not be optimal since they are usually trained by general images. Thus, we propose the adaptive deep PnP algorithm for every testing data, which can online update the weights and bias in networks adaptively according to the special dynamic scene. Besides, aiming at the color video SCI, we use the proposed deep demosaicing method to improve the reconstruction quality.  

\subsection{Online PnP}
In general, every dynamic scene has its own special features which may not be included in the training dataset. 
Towards this end, we optimize the deep denosing prior by online updating the network according to the special test dynamic scene. As shown in Fig.~\ref{fig:principle} (b), the network, specifically, the FFDNet and FastDVDNet, in the {\em adaptive deep denoising prior} is updated by minimizing the following loss during the PnP iterations:
\begin{equation}
	\ell = \|\yv - \Hmat \Tmat_{\cal M}\xv^{(k)}\|_2^{2},
	\label{Eq:loss}
\end{equation}
due to the fact that the output of the adaptive deep denoising prior represents the dynamic scene $\xv$ and $\xv^{(k)}$ approaches closer to the true dynamic scene $\xv$ when the loss $\ell$ in (\ref{Eq:loss}) decreases. 

To be concrete, in the ${k}$-th, iteration, as shown in Fig.~\ref{fig:principle}, the projection is first computed based on input containing the measurement $\yv$, the matrix ${\Hmat}$ and the $\tilde{\xv}^{k-1}$ output of the mosaicing process from $\{\Xmat\}$ to $\{\tilde\Xmat\}$. Then the interleaving and demosacing are conducted, where the size of data becomes into ${{n_x}\times {n_y}\times 3\times B}$ from ${\frac{n_x}{2}\times \frac{n_y}{2} \times B \times 4}$. Afterwards, the adaptive deep denoising prior denoises the output images. Here, the output of adaptive deep denoising prior can be regarded as the last reconstructed result when this computation is the last iteration in PnP, while the output should be mosaiced when this is not the last iteration. In the gray case, the interleaving, demosacing and mosacing processes are not necessary to carry out. Because the initial iterations in PnP framework just compute the rough features, we did not update the network in denoising prior at the first few iterations. To avoid the over-fitting phenomenon in this self-supervised learning process~\cite{ulyanov2018dip}, we update the network once every several iterations. 

In summary, the online PnP adds one more step, \ie, to update the network parameters in the deep denoiser, every $K_0$ iterations, leading to the proposed adaptive PnP algorithm for color video SCI exhibited in Algorithm~\ref{algo:Adaptive_PnP_Color_Video}.

\begin{algorithm}[!htbp]
	\caption{Adaptive PnP for Color Video SCI}
	\begin{algorithmic}[1]
		\REQUIRE $\Hmat$, $\yv$, $\Tmat_{\cal M}$.
		\STATE Initial $\xv,\uv,\vv,\wv, \rho, \tau$ and $K_0$.
		\FOR{$k=0$ to $K_{\rm max}$}
		\STATE Update $\qv^{(k+1)}$ by Eq.~\eqref{Eq:qv_k+1}. 
		\STATE Update $\xv^{(k+1)}$ by Eq.~\eqref{Eq:xv_k+1} or by the deep demosaicing network \eqref{Eq:deep_dema}.
		\STATE Update $\vv^{(k+1)}$ by deep denoiser in Eq.~\eqref{Eq:vvk+1}.
		\STATE Update $\wv^{(k+1)}$ by Eq.~\eqref{Eq:wvk+1}.
		\STATE Update $\uv^{(k+1)}$ by Eq.~\eqref{Eq:uv_k+1}.
		\IF{{\rm mod}($k, K_0$) =0 AND $k>0$}
		    \STATE Update network parameters in the deep prior.
		\ENDIF
		\ENDFOR
	\end{algorithmic}
	\label{algo:Adaptive_PnP_Color_Video}
\end{algorithm}

\begin{table*}[!htbp]
	\caption{Mid-scale Bayer benchmark dataset: the average results of PSNR in dB (left entry in each cell) and SSIM (right entry) and running time (all set 80 iterations for more equitable comparison) per measurement/shot in minutes by different algorithms on 6 benchmark color Bayer datasets used in~\cite{PnP_SCI_TPAMI2021}.} 
	\centering
	\resizebox{1\textwidth}{!}
	{
\begin{tabular}{l|llllll|ll} 
\toprule
Algorithm                                                                             & Beauty                 & Bosphorus              & Jockey                 & Runner                 & ShakeNDry              & Traffic                & Average                & Run~time (min)      \\ 
\hline
GAP-TV~\cite{Yuan16ICIP_GAP_y}                                                                                & 33.08,  0.9639         & 29.70,  0.9144         & 29.48,  0.8874         & 29.10,  0.8780         & 29.59,  0.8928         & 19.84,  0.6448         & 28.47, 0.8636          & \textbf{0.3 (CPU)}  \\
DeSCI~\cite{Liu18TPAMI_y}                                                                                 & 34.66, 0.9711          & 32.88, 0.9518          & 34.14, 0.9382          & 36.16, 0.9489          & 30.94, 0.9049          & 24.62, 0.8387          & 32.23, 0.9256          & 1544 (CPU)          \\
\hline
GAP-FFDNet~\cite{Yuan_2020_CVPR_y}                                                                            & 33.88,  0.9665         & 33.13,  0.9540         & 34.25,  0.9350         & 34.67,  0.9271         & 32.24,  0.9379         & 24.02,  0.8299         & 32.03, 0.9251          & 0.95 (CPU+GPU)      \\
Two-Stage ADMM-FFDNet                                                                     & 35.25,~0.9724          & 32.62,~0.9510          & 34.95,~0.9473          & 35.16,~0.9404          & 32.13,~0.9303          & 24.58,~0.8211          & 32.45,~0.9271          & 0.34 (GPU)          \\
Two-Stage ADMM-FFDNet (online)                                                            & 35.39,~0.9729          & 33.11,~0.9517          & 35.10,~0.9486          & 35.41,~0.9422          & 32.30,~0.9338          & 24.87,~0.8217          & 32.70,~0.9285          & 0.35 (GPU)          \\
Two-Stage ADMM-FFDNet-CFD                                                                 & 33.29,~0.9653          & 29.98,~0.9055          & 28.83,~0.8741          & 30.31,~0.8871          & 29.45,~0.8758          & 20.45,~0.6720          & 28.72,~0.8633          & 0.34 (GPU)          \\
Two-Stage ADMM-FFDNet-DDNet                                                               & 35.30,~0.9723          & 33.42,~0.9549          & 34.85,~0.9454          & 35.44,~0.9422          & 32.60,~0.9317          & 25.19,~0.8252          & 32.83,~0.9286          & 0.62 (GPU)          \\
Two-Stage ADMM-FFDNet (online)-DDNet                                                      & 35.38, 0.9727          & 33.54, 0.9549          & 35.07, 0.9479          & 35.40, 0.9424          & 32.73, 0.9346          & 25.21, 0.8262          & 32.89, 0.9298          & 0.63 (GPU)          \\
\rowcolor[rgb]{0.753,0.753,0.753} Two-Stage ADMM-FFDNet (online-reuse)-DDNet              & 35.45,~0.9730          & 33.62,~0.9550          & 35.16,~0.9496          & 35.37,~0.9435          & 32.83,~0.9367          & 25.25,~0.8277          & 32.95,~0.9309          & 0.63 (GPU)          \\
\hline
GAP-FastDVDNet~\cite{PnP_SCI_TPAMI2021}                                                                        & 35.12,  0.9709         & 35.80,  0.9701         & 35.14,  0.9442         & 38.00,  0.9618         & 33.35,  0.9452         & 27.22,  0.9085         & 34.11, 0.9501          & 0.95 (CPU+GPU)      \\
Two-Stage ADMM-FastDVDNet-DDNet                                                           & \textbf{36.24,~0.9752} & \textbf{36.56,~0.9726} & 36.34,~0.9558          & \textbf{38.00},~0.9592 & 34.27,~0.9451          & \textbf{28.15,~0.9104} & \textbf{34.93,~0.9530} & 0.69 (GPU)          \\
\rowcolor[rgb]{0.761,0.757,0.757}  Two-Stage ADMM-FastDVDNet (online-reuse)-DDNet & 36.22,~\textbf{0.9752}          & 36.51,~0.9724          & \textbf{36.36,~0.9560} & 37.85,~\textbf{0.9593} & \textbf{34.31,~0.9456} & \textbf{28.15,~0.9104} & 34.90, \textbf{0.9531}          & 0.76 (GPU)          \\
\bottomrule
\end{tabular}
	}
	\label{Tab:results_midscale}
\end{table*}
\begin{figure*}[htbp!] 
\centering
\includegraphics[width=1\linewidth]{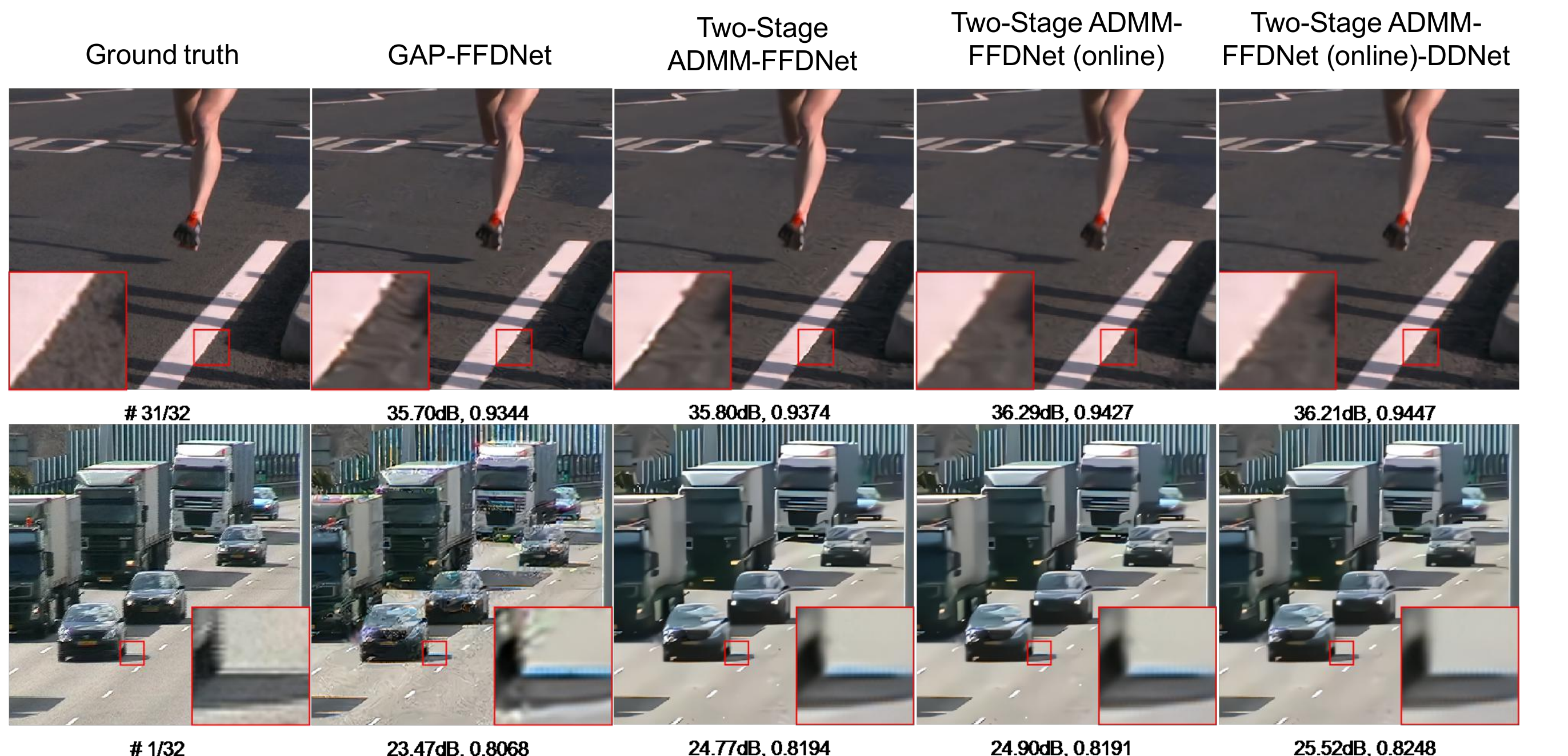}
\vspace{-4mm}
\caption{Selected reconstruction result comparison of mid-scale color Bayer benchmark by GAP-FFDNet, Two-Stage ADMM-FFDNet, Two-Stage ADMM-FFDNet (online) and Two-Stage ADMM-FFDNet (online)-DDNet.
}
\label{fig:midscale_sim_ablation}
\end{figure*}

\begin{figure*}[htbp!] 
\centering
\includegraphics[width=1\linewidth]{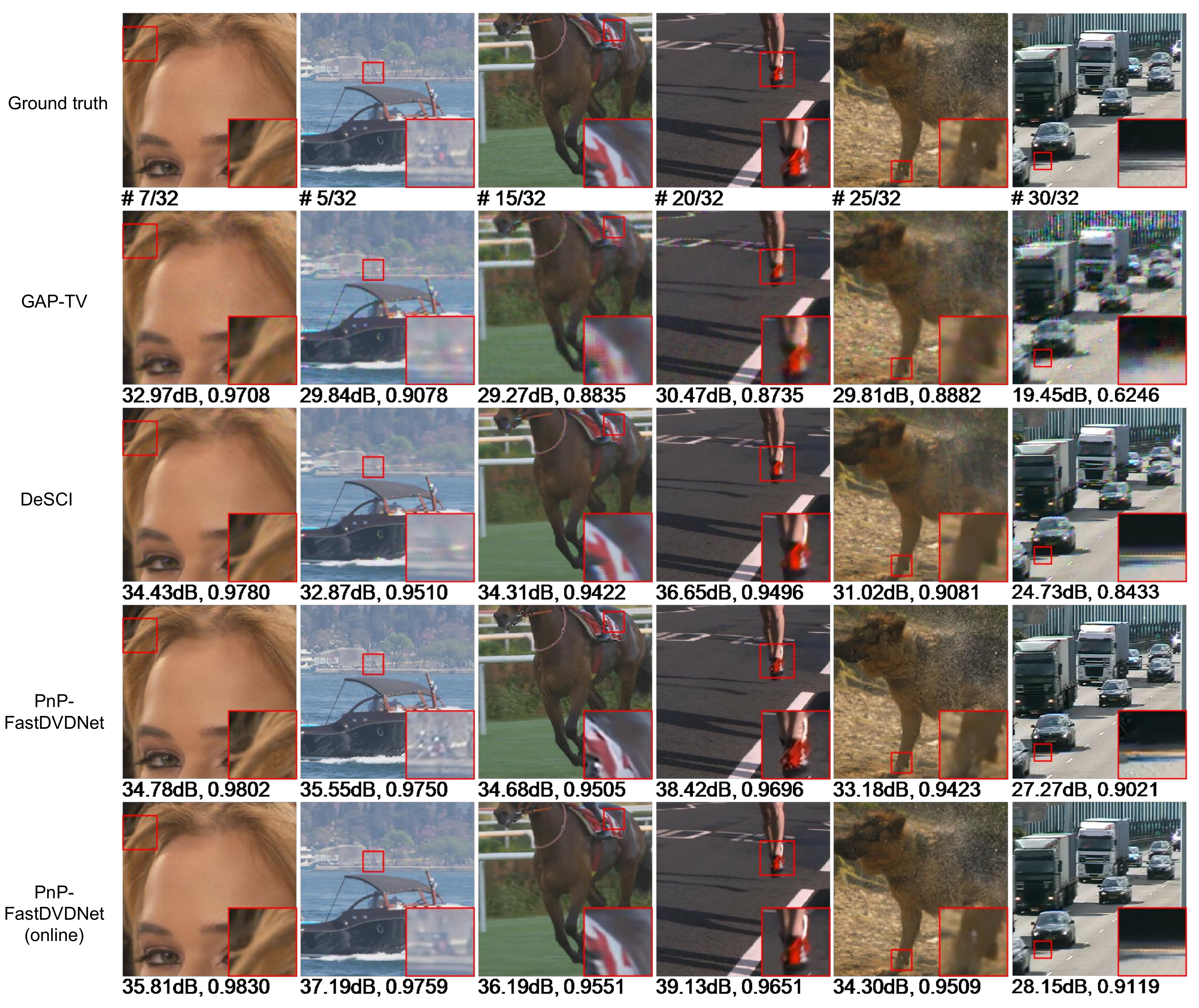}
\vspace{-4mm}
\caption{Selected reconstruction results of mid-scale color Bayer benchmark by GAP-TV~\cite{Yuan16ICIP_GAP_y}, DeSCI~\cite{Liu18TPAMI_y}, PnP-FastDVDNet~\cite{PnP_SCI_TPAMI2021} and the proposed PnP-FastDVDNet (online).
}
\label{fig:midscale_sim}
\end{figure*}

\subsection{Online PnP for Continuous Measurements}
To further speed up the adaptation process, consider the multiple sequential measurements for the scene, \ie, the background is barely changed. 
In this case, after the first measurement, the deep denoiser should be getting close to the desired video.
Therefore, for the second measurement, instead of initializing the deep denoiser using the pre-trained parameters, we can directly use the updated, thus adapted, network parameters from the first measurement. 
Following this, the third measurement can use the parameters updated from the second measurements, and so on and so forth. 
We can also decrease the frequency of updating the network parameters for the latter measurements, \ie, by increasing $K_0$ in Algorithm~\ref{algo:Adaptive_PnP_Color_Video}.


\section{Results \label{Sec:Results}}
To validate the proposed adaptive deep PnP algorithm, we apply the proposed PnP algorithms to both simulation~\cite{Liu18TPAMI_y,Ma19ICCV_y} and real datasets captured by the SCI cameras~\cite{Patrick13OE_y,2014CVPR_xin_y,Qiao2020_APLP_y}. Conventional denoising algorithms including TV~\cite{Yuan16ICIP_GAP_y}, and DeSCI ~\cite{Liu18TPAMI_y} are used for comparison. For the deep learning based denoiser, we adopt FFDNet~\cite{Zhang18TIP_FFDNet_y} and FastDVDNet~\cite{Tassano_2020_CVPR_y}.
Once the adaptive method adopted on the deep learning based denoiser, the reconstruction quality can improve. Both PSNR and SSIM~\cite{Wang04imagequality_y} are employed as metrics to compare different algorithms.

\subsection{Benchmark Data: Color RGB-Bayer Videos \label{sec:sim_color}}
In this section, we adopt the mid-scale color RGB video testing dataset from~\cite{PnP_SCI_TPAMI2021}, which has 6 scenes of spatial size $512\times512\times3$, where $3$ denotes the RGB channels. The compression rate is set to $B=8$. We performed mosaicing , interleaving and demosacing, as shown in Fig.~\ref{fig:principle}, which are not necessary in the grayscale case. For each dataset, we have 4 compressed measurements and thus in total 32 RGB video frames.
As shown in Fig.~\ref{fig:midscale_sim}, these datasets include \texttt{Beauty}, \texttt{Bosphorus}, \texttt{Jockey}, \texttt{ShakeNDry}, \texttt{Runner}
and \texttt{Traffic}. 

To verify the performance of our proposed two-stage ADMM, adaptive deep PnP algorithm (online) as well as the deep demosaicing method, we have compared the original PnP-FFDNet-color (named GAP-FFDNet in Table~\ref{Tab:results_midscale}) with Two-Stage PnP-ADMM-FFDNet-color (named Two-Stage ADMM-FFDNet), compared Two-Stage PnP-ADMM-FFDNet-color with Two-Stage Adaptive PnP-ADMM-FFDNet-color (named Two-Stage ADMM-FFDNet (online)) and Two-Stage Adaptive PnP-ADMM-FFDNet-color deep demosaicing (named Two-Stage ADMM-FFDNet (online)-DDNet).  Here, `Malvar04'~\cite{malvar2004high-quality_y} method is adopted to demosaic in the GAP-FFDNet, GAP-FastDVDNet,  Two-Stage ADMM-FFDNet, Two-Stage ADMM-FFDNet (online). 

Fig.~\ref{fig:midscale_sim_ablation} shows the comparison of each module used in our online PnP. The \texttt{Runner} video in the upper part shows that both Two-Stage ADMM and adaptive strategy for PnP provide smoother results. Meanwhile, the \texttt{Traffic} video in the lower part indicates our video deep demosaicing network (DDNet) recover the color more accurately in the reconstruction.

For PnP-FastDVDNet, the Two-Stage ADMM framework also achieves higher quality, but the adaptive PnP show little differences with the non-adaptive ones. This may be because the color FastDVDNet model has trained to a good point and any tiny variance can influence the high denoising quality. 
Table~\ref{Tab:results_midscale} summarizes the PSNR and SSIM results of these datasets. 
We have the following observations:
\begin{itemize}
    \item [$i)$] On average, the proposed Two-Stage  ADMM optimization  can obtain 0.42dB higher PSNR than GAP using FFDNet-color as the denoiser in PnP. 
    \item [$i)$] Our proposed online strategy leads to a 0.15dB (from 32.45 to 32.70) improvement when using FFDNet-color as the denoiser. 
    \item [$ii)$] By adopting the deep demosaicing method, Two-Stage ADMM-FFDNet (online)-DDNet achieves better results than Two-Stage ADMM-FFDNet (online), i.e., more than 0.19dB in PSNR and 0.0013 in SSIM.
    \item [$iv)$] Again, it is easy to achieve better results with online in deep priors, \ie, PnP-FastDVDNet(deep demosaic), PnP-FastDVDNet and PnP-FFDNet, than their counterparts without online. However, the improved quality is limited by the network. 
    
    \item [$v)$] Due to adopting the proposed adaptive deep prior (online) and deep demosaicing method, the PnP-FastDVDNet-color (online+deep demosaic) achieves the best results among all these algorithms, setting a new state-of-the-art.

    \item [$vi)$] In this mid-scale color case, all these six deep prior methods in the lower part of Table~\ref{Tab:results_midscale}
    outperform DeSCI. Besides, the deep prior methods just need about 1.6 minutes while DeSCI demands 1544 minutes.
\end{itemize}

\begin{figure}[htbp!] 
\vspace{-2mm}
\centering
\includegraphics[width=1\linewidth]{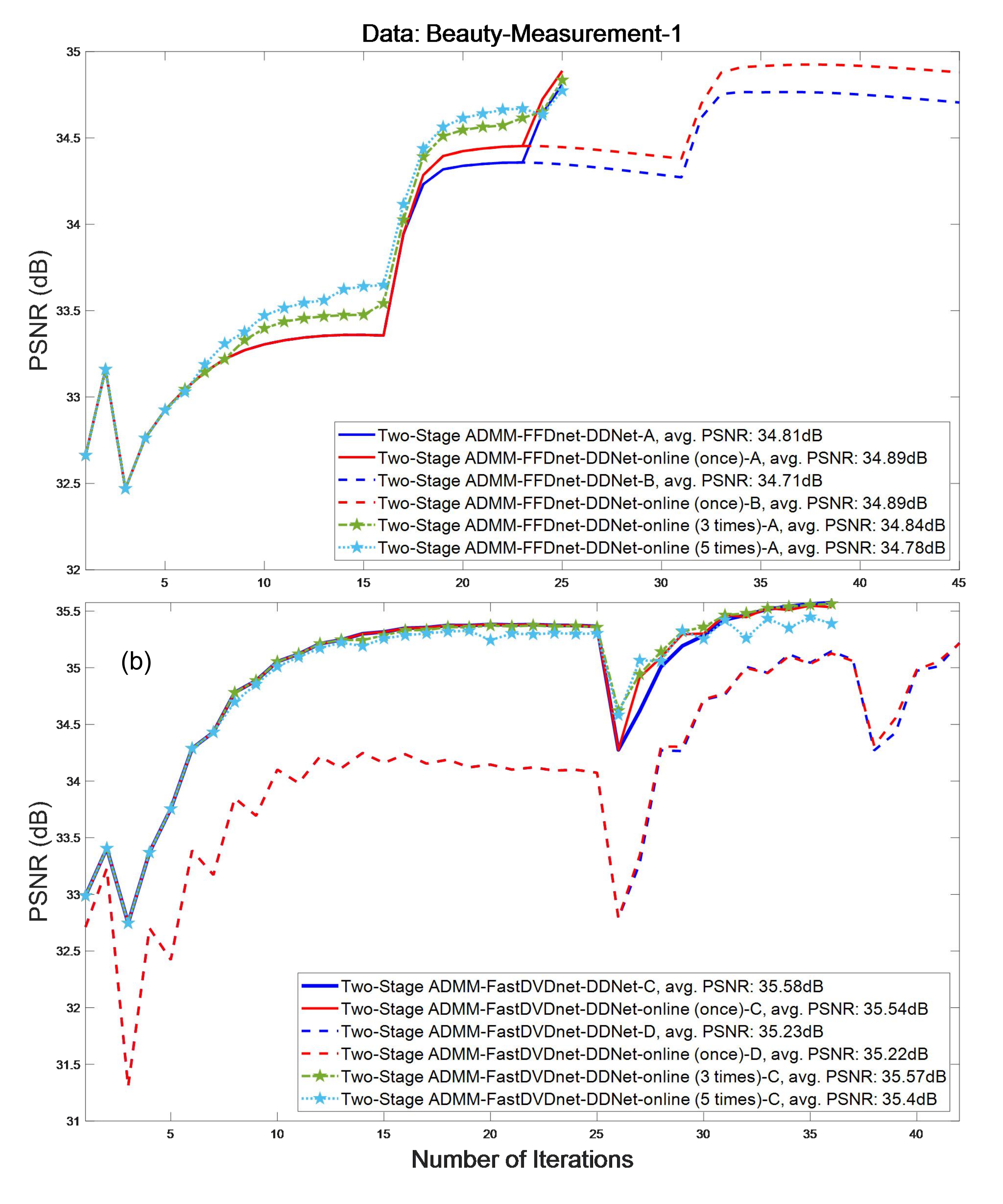}
\vspace{-3mm}
\caption{The comparisons of different parameter settings using FFDNet (a) and FastDVDNet (b) as denoiser. The capital letter in the end of each algorithms means different parameter settings. \\
A: $\sigma$=25 (15 iter), $\sigma$=12 (7 iter), $\sigma$=6 (3 iter). \\
B: $\sigma$=25 (15 iter), $\sigma$=12 (15 iter), $\sigma$=6 (15 iter).\\
C: $\sigma$=12 (24 iter), $\sigma$=6 (12 iter).\\
D: $\sigma$=25 (24 iter), $\sigma$=12 (12 iter), $\sigma$=6 (6 iter).\\
}
\label{fig:compare_params}
\end{figure}

\begin{figure*}[htbp!] 
	\begin{center}
\includegraphics[width=1\linewidth]{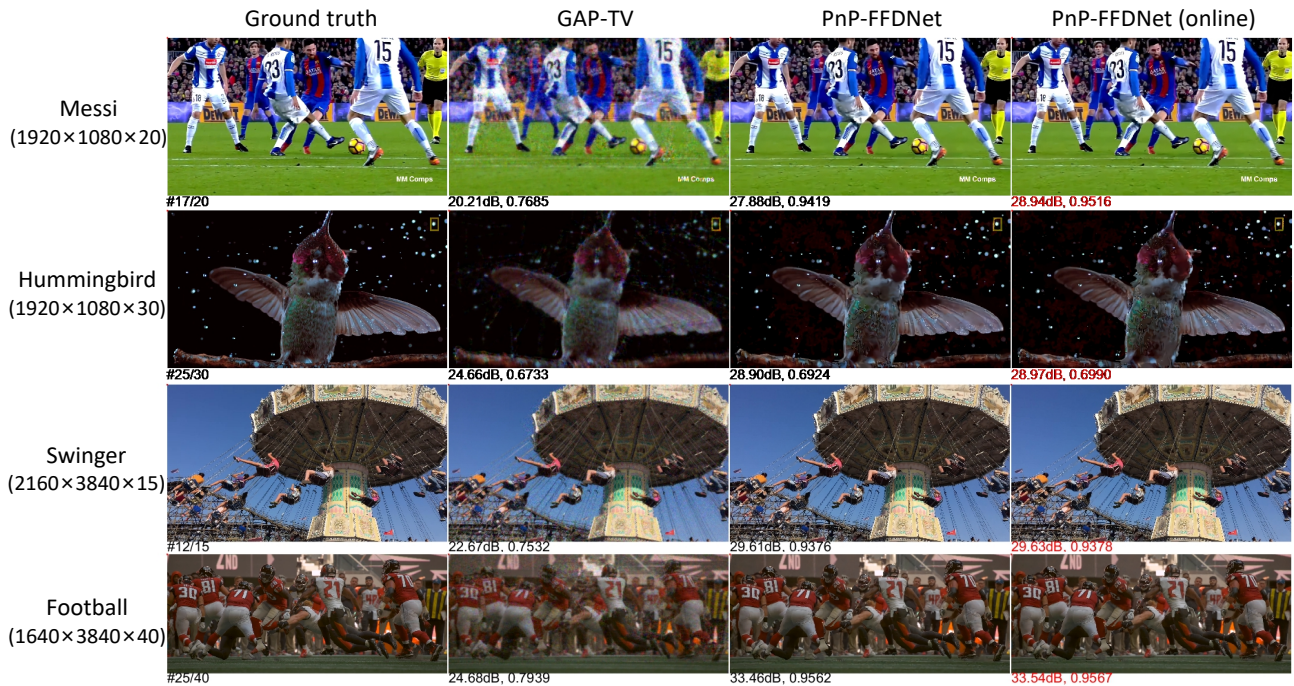}
	\end{center}
\centering
\caption{Selected reconstruction results of large-scale color videos by GAP-TV~\cite{Yuan16ICIP_GAP_y}, PnP-FFDNet~\cite{Yuan_2020_CVPR_y} and proposed PnP-FFDNet (online). One frame is plotted for each scene and the PSNR and SSIM values are shown in the bottom-left.
}
\label{fig:largescale_sim}
\end{figure*}

Fig.~\ref{fig:midscale_sim} plots selected reconstruction frames of different algorithms for these 6 RGB Bayer datasets, where we only plot against GAP-TV, DeSCI and PnP-FastDVDNet (color), which is the previous state-of-the-art. The Two-Stage ADMM-FastDVDNet (online-reuse)-DDNet here is briefly named as PnP-FastDVDNet(online).  
It can be seen from Fig.~\ref{fig:midscale_sim} that GAP-FastDVDNet (named PnP-FastDVDNet) and Two-Stage ADMM-FastDVDNet (online-reuse)-DDNet (named PnP-FastDVDNet (online)) are providing smooth motions and fewer artifacts. Some color mismatch exists in the GAP-TV, DeSCI. For instance, in the {\texttt{Traffic}} data, the color of the cars is incorrectly reconstructed for these methods; a similar case exists in the water drops in the {\texttt{ShakeNDry}}. Overall, GAP-TV provides blurry results and DeSCI sometimes over-smooths the background such as the lawn in the {\texttt{Jockey}} data. 
Compared with the previous state-of-the-art, PnP-FastDVDNet, our proposed online method is providing fewer twists and artifacts.
For example, the lines of the zoomed cars in the {\texttt{Traffic}} frame in Fig.~\ref{fig:midscale_sim}, shoes in the \texttt{runner}.
For some specific reconstructed frames, the benefits from our adaptive method are more than 2 dB in PSNR, \eg, the frames shown in {\texttt{Jockey}}. 

\subsection{Hyper-parameter Setting \label{Sec:params}}
Same as the PnP framework, we need to set hyper-parameters like noise level $\sigma$ and the number of iterations. In Fig.~\ref{fig:compare_params}, we choose the first measurement in data \texttt{Beauty} as an example. All algorithms are initialized with TV prior as their warm starting points. The learning rate of online algorithms is set $1\times10^{-6}$. 

For the denoiser of FFDNet (the upper plot in Fig.~\ref{fig:compare_params}), unlike the parameter settings in \cite{PnP_SCI_TPAMI2021} (B), in which the iterations for each $\sigma$ are the same, our Two-Stage ADMM-FFDNet-DDNet halves the iterations for smaller $\sigma$ (A), and requires fewer iterations to achieve similar results. Therefore it can reduce a significant amount of convergence time. For adaptive cases (online), we test the running time of updating networks in one measurement. These results prove that updating the network once is enough for the network, and can reach higher PSNR than 3 times or 5 times updating. This mayd due to the strong learning capability of FFDNet.

For the denoiser FastDVDNet (the lower plot in Fig.~\ref{fig:compare_params}), parameter settings are kind of different from FFDNet ones. The initial $\sigma$ is set to 12 rather than 25 (in FFDNet), but we get better results and need fewer iterations. The adaptive cases are the same as those used in FFDNet.

These observations tell us we can use early stopping in a certain iteration to reach the best results and do not need much network updating for one measurement. Therefore, though parameter settings for each data are not identical, we can follow these observations to reduce the final convergence time. {That also means the time to reach the performance shown in Table~\ref{Tab:results_midscale} can be much less than the running time for comparison there (some saves more than half running time)}. 

\subsection{Large-scale Data Results}
We have also tried our proposed online method on large-scale datasets used in~\cite{PnP_SCI_TPAMI2021}. 
Here, the simulated color video SCI measurements for large-scale data is from four YouTube slow-motion videos, more details can be referred to ~\cite{PnP_SCI_TPAMI2021}. 
Because of the limited memory of GPU, we only utilized GAP-TV, GAP-FFDNet (named PnP-FFDNet) and Two-Stage ADMM-FFDNet (online)-DDNet (named PnP-FFDNet(online)). 

From Fig~\ref{fig:largescale_sim}, we can see that $i$) because of existences of many fine details, GAP-TV cannot provide high quality results; $ii$) both PnP-FFDNet and PnP-FFDNet (online) lead to significant improvements over GAP-TV (at least 4.24 dB in PSNR), and $iii$) PnP-FFDNet (online)  leads to best results on the four datasets and better than that by the PnP-FFDNet. In short, large-scale datasets further verified the flexibility and superior performance of our proposed adaptive deep prior pnp method. 

\subsection{Benchmark Data: Grayscale Videos \label{sec:sim_gray}}
We follow the simulation setup in~\cite{Yuan_2020_CVPR_y} of six datasets, \ie, \texttt{Kobe, Traffic, Runner, Drop, crash,} and \texttt{aerial}, where $B=8$ video frames are compressed into a single measurement.
Table~\ref{Tab:results_4video} summarizes the PSNR and SSIM results of these 6 benchmark data using various denoising algorithms, where DeSCI can be categorized as GAP-WNNM, and GAP-FastDVD (online-reuse) uses the proposed adaptive method in deep denoising prior while  GAP-FastDVDNet does not.  Because these datasets represent different dynamic scenes, we chose different parameters for different datasets. Correspondingly,
the learning rate is also fine tuned for different datasets, as well as the time of interval iterations between two online updating networks. The same strategies are also used in color RGB video  and real datasets.
It can be observed from Table~\ref{Tab:results_4video} that: 
\begin{itemize}
    \item [$i$)] 
    Our proposed online strategy leads to a 0.21dB (from 32.35 to 32.56) improvement when using FastDVDNet as the denoiser in PnP. 
	\item [$ii$)] 
	GAP-FastDVDNet (online) can achieve similar results with DeSCI, less than 0.11dB in PSNR but more than 0.0093 in SSIM, but with a much shorter running time. Among these fast algorithms, namely, GAP-TV, GAP-FastDVDNet and GAP-FastDVDNet (online), our proposed GAP-FastDVDNet (online) achieves the best result. 
\end{itemize}

\begin{figure*}[htbp!] 
\centering
\includegraphics[width=.9\textwidth]{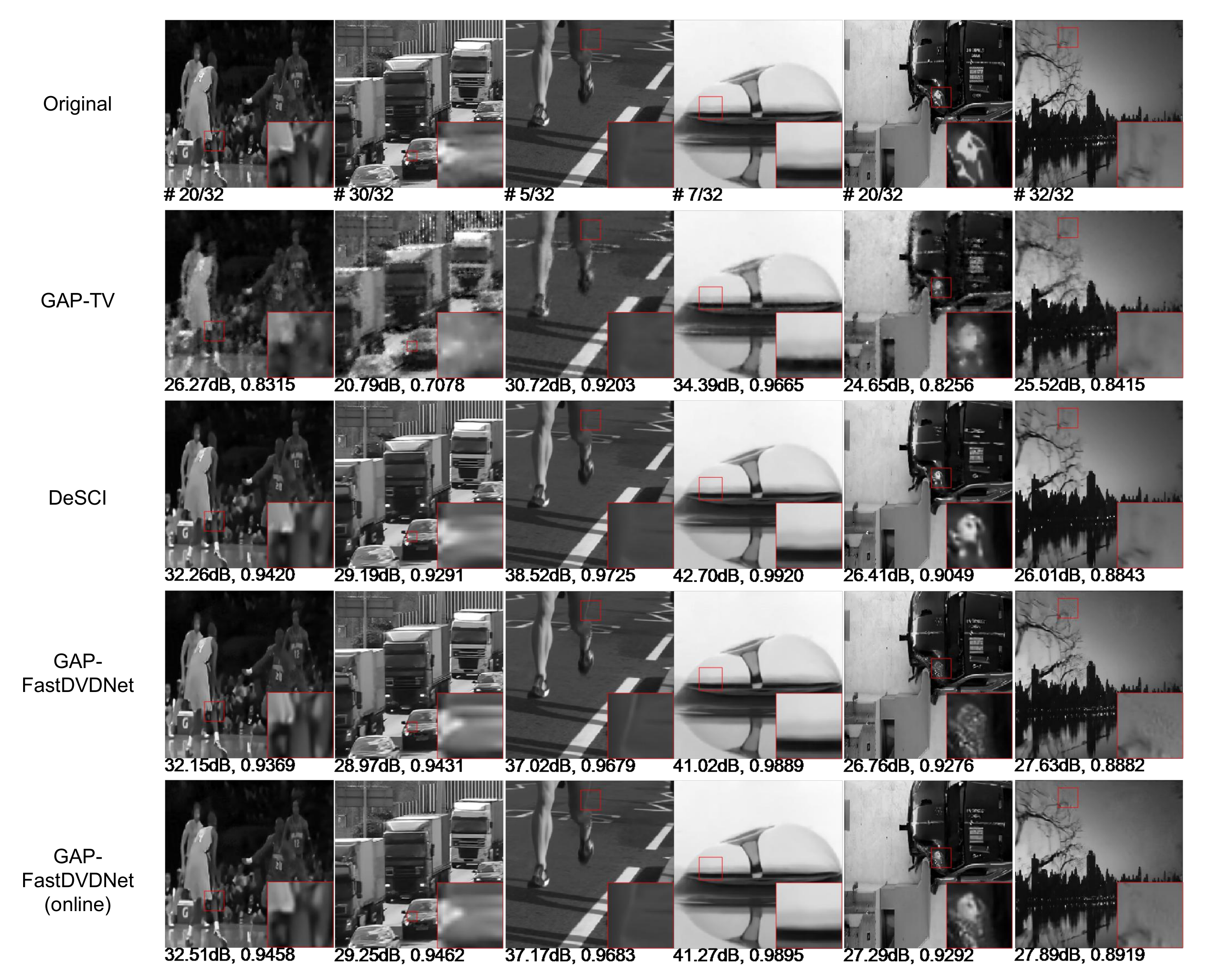}
\caption{Selected reconstruction results of gray-scale benchmark by GAP-TV~\cite{Yuan16ICIP_GAP_y}, DeSCI~\cite{Liu18TPAMI_y},  GAP-FastDVDNet~\cite{PnP_SCI_TPAMI2021} and proposed GAP-FastDVDNet (online).
}
\label{fig:greyscale_sim}
\end{figure*}

Fig.~\ref{fig:greyscale_sim} plots selected frames of these six datasets using different algorithms. It can be seen that GAP-FastDVDNet (online) can achieve better results than the other five algorithms including DeSCI, especially in some fine details, as shown in the zoomed parts of \texttt{Aerial} in Fig.~\ref{fig:greyscale_sim}. Even though DeSCI still leads to the highest average PSNR, the difference between gap-fastDVDNet (online) and DeSCI is very small. What's more,  gap-fastDVDNet (online) has a higher average SSIM than those of DeSCI, and can also provide finer details. Besides, gap-fastDVDNet (online) has a faster speed than DeSCI. 

\begin{table*}[!htbp]
	\caption{Grayscale benchmark dataset: the average results of PSNR in dB (left entry in each cell) and SSIM (right entry) and running time (all set 80 iterations for more equitable comparison) per measurement/shot in minutes by different algorithms on 6 benchmark datasets.}
	\centering
	\resizebox{\textwidth}{!}
	{

	\begin{tabular}{l|llllll|ll} 
\toprule
Algorithm                                                       & Kobe                   & Traffic                & Runner                 & Drop                   & Crash                  & Aerial                 & Average                & Running time (mins)      \\ 
\hline
GAP-TV~\cite{Yuan16ICIP_GAP_y}                                                           & 26.92, 0.8378          & 20.66, 0.6905          & 29.81, 0.8949          & 34.95, 0.9664          & 24.48, 0.7988          & 24.81, 0.8105          & 26.94, 0.833           & \textbf{0.03 (CPU)}  \\
DeSCI~\cite{Liu18TPAMI_y}                                                           & \textbf{33.25, 0.9518} & \textbf{28.71, 0.9250} & \textbf{38.48, 0.9693} & \textbf{43.10, 0.9925} & 27.04, 0.9094          & 25.33, 0.8603          & \textbf{32.65}, 0.9347 & 103 (CPU)            \\
GAP-FastDVDNet~\cite{PnP_SCI_TPAMI2021}                                                 & 32.73, 0.9466          & 27.95, 0.9321          & 36.29, 0.9619          & 41.82, 0.9892          & 27.32, 0.9253          & 27.98, 0.8966          & 32.35, 0.9420          & 0.1 (CPU+GPU)        \\
\rowcolor[rgb]{0.757,0.757,0.757} GAP-FastDVDNet (online-reuse) & 32.95, 0.9514          & 28.16, 0.9349          & 36.41, 0.9622          & 41.95, 0.9899          & \textbf{27.64, 0.9288} & \textbf{28.24, 0.8974} & 32.56, \textbf{0.9441}          & 0.2 (CPU+GPU)        \\
\bottomrule
\end{tabular}

	}
	\label{Tab:results_4video}
\end{table*}

Therefore, both the grayscale and color video datasets have verified the superiority of our proposed adaptive PnP framework and next we apply it to the real data captured by SCI cameras.   

\begin{figure}[htbp!] 
\centering
\includegraphics[width=1\linewidth]{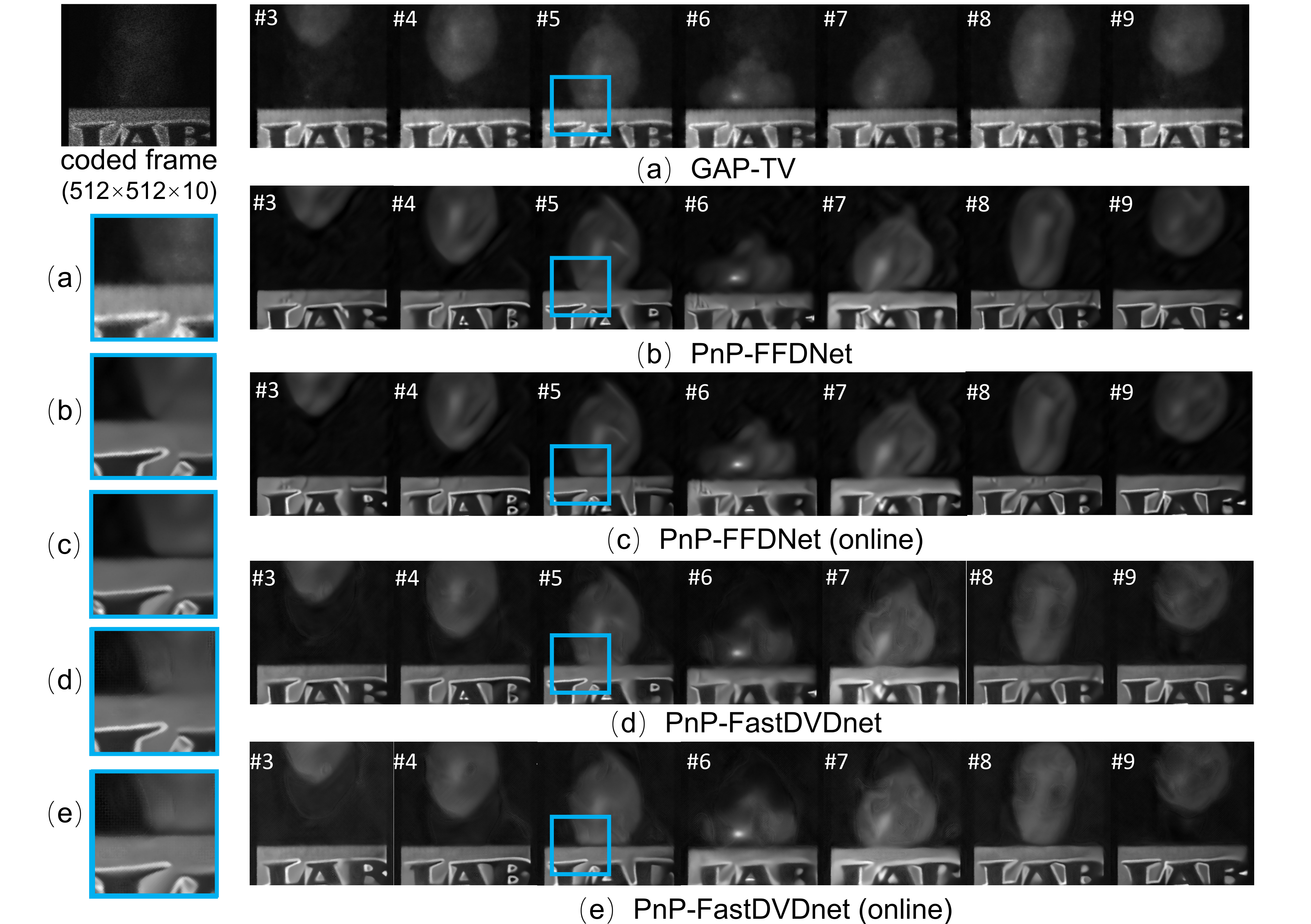}
\vspace{-3mm}
\caption{Proposed online PnP on the grayscale real data compared with others. A $512\times512\times10$ video reconstructed from a snapshot measurement; 7 out of 10 frames are shown here.
}
\label{fig:real_gray}
\end{figure}

\subsection{Real Data \label{Sec:realdata}}
Lastly, we apply the proposed adaptive deep PnP framework to real datasets captured by gray~\cite{Patrick13OE_y,Qiao2020_APLP_y} and color~\cite{2014CVPR_xin_y} SCI cameras respectively, to verify the robustness of the proposed algorithm. 

In the grayscale case, Fig.~\ref{fig:real_gray} plots 7 out of 10 reconstructed frames of the \texttt{waterball}, which shows the dropping and rebounding process of the waterball within 20 milliseconds~\cite{Qiao2020_APLP_y}. Similar to the simulation, we apply our online PnP to both FFDNet and FastDVDNet. It can be seen that both methods outperform their counterparts without an online strategy. Specifically, our online methods in Fig.~\ref{fig:real_gray}(c) and (e) show sharp boundaries and fine details. Please refer to the zoomed parts shown on the left of Fig.~\ref{fig:real_gray}.

\begin{figure}[htbp!] 
\centering
\includegraphics[width=1\linewidth]{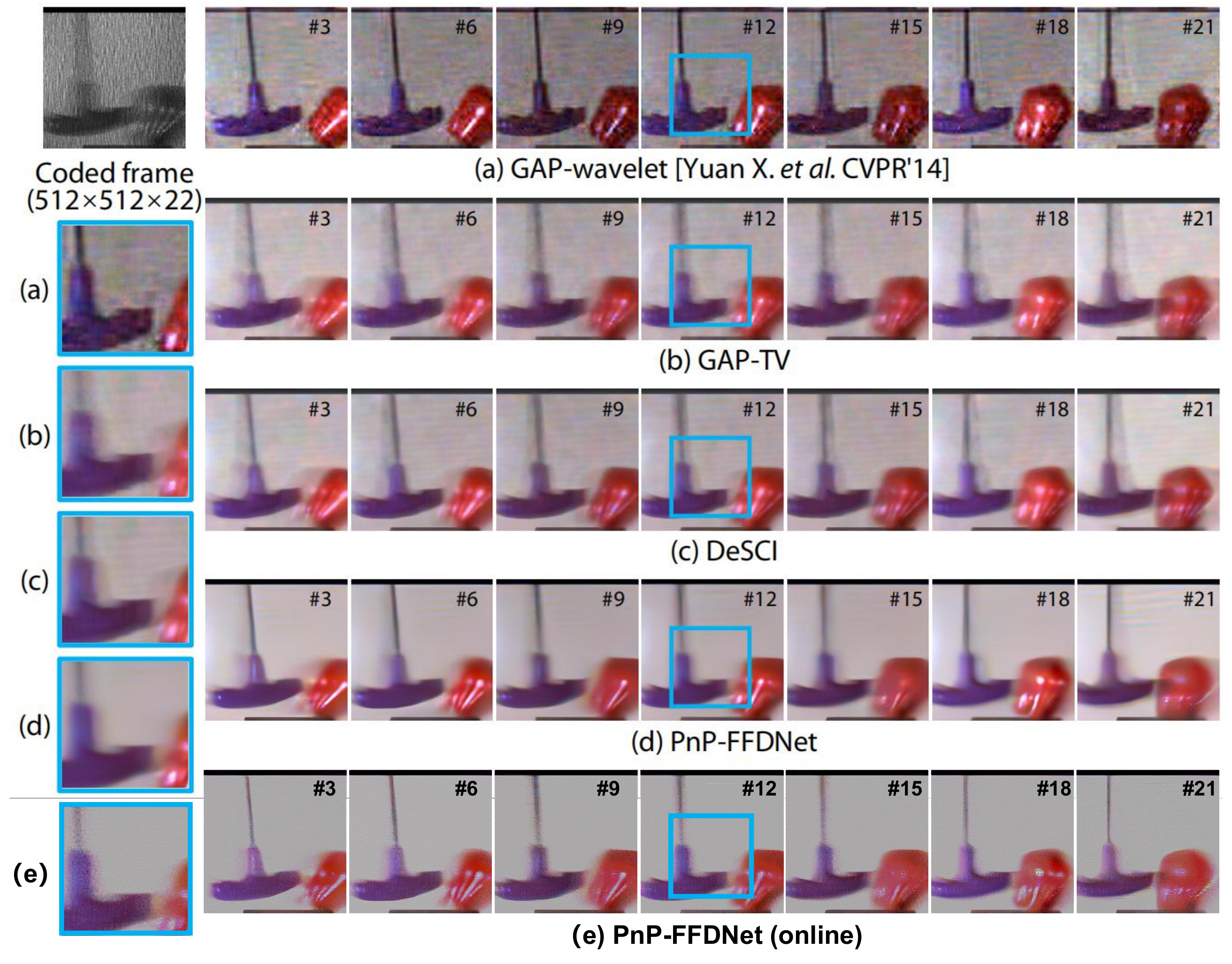}
\vspace{-3mm}
\caption{Proposed online PnP-FFDNet on the color real data compared with others. A $512\times512\times 3\times22$ RGB color video reconstructed from a snapshot Bayer measurement; 7 out of 22 frames are shown here.
}
\label{fig:real_color}
\end{figure}

In the color case, following the procedure in the mid-scale color data, an RGB video of $B=22$ frames with size of $512\times 512 \times 3$ is reconstructed from a single Bayer mosaic measurement shown in Fig.~\ref{fig:real_color} of the data \texttt{hammer}~\cite{2014CVPR_xin_y}. It can be seen that there are more noises in the results achieved by  GAP-wavelet~\cite{2014CVPR_xin_y} and some blurry phenomenon in GAP-TV.
PnP-FFDNet shows sharper edges than DeSCI with a clean background. By  adaptive updating the network, PnP-FFDNet (online) removes more background noises and provides clearer images than its counterpart without online updating.
PnP-FastDVDNet (online) can also improve the result of PnP-FastDVD but the visual quality is not as good as PnP-FFDNet and thus omitted in this plot. 

These real data results clearly verify the feasibility of our proposed adaptive deep PnP algorithm in real SCI cameras. 



\section{Conclusions \label{Sec:conclu}}
We have proposed an adaptive plug-and-play framework for video snapshot compressive imaging reconstruction.
By updating the parameters of the deep denoising network online with the PnP iterations, better results can be achieved due to the fact that the deep denoiser is adapted to the specific video being reconstructed. 
A new two-stage ADMM optimization and deep video demosaicing network have also been proposed to further improve the reconstruction quality of color SCI videos. 
Both simulation and real data results verified the outstanding performance of the proposed adaptive PnP algorithm.

It has always been a trade-off in speed, adaptivity and accuracy for the reconstruction in computational imaging. Our proposed online method provides a feasible solution to fine-tune the pre-trained denoising network during reconstruction by paying the price of affordable additional running time. Our proposed semi-supervised online strategy can also be used in  end-to-end networks for SCI reconstruction. 

In addition, our algorithm still has space to improve performance and speed, because our deep denoising prior and demosaicing prior can be jointly trained like other recent restoration algorithms~\cite{chi2019joint,a2021beyond_joint,guo2021joint} to make it more adaptive and faster for video SCI. Our future work will focus on the joint prior in the semi-supervised PnP framework.

\bibliographystyle{spmpsci}      
\bibliography{bibliography}


\end{document}